\documentclass[11pt,twoside]{article}
\usepackage{asp2010}

\resetcounters
\begin{document}

\title{Prediction of close binarity based on planetary nebula morphology}
\author{B. Miszalski$^1$, R.L.M. Corradi$^{2,3}$, D. Jones$^{4}$, M. Santander-Garc\'ia$^{5,2,3}$,\\ P. Rodr\'iguez-Gil$^{5,2,3}$ and M. M. Rubio-D\'iez$^{5,6}$
\affil{$^1$Centre for Astrophysics Research, STRI, University of Hertfordshire, College Lane Campus, Hatfield AL10 9AB, UK}
\affil{$^2$Instituto de Astrof\'isica de Canarias, E-38200 La Laguna, Tenerife, Spain}
\affil{$^3$Departamento de Astrof\'isica, Universidad de La Laguna, E-38205 La Laguna, Tenerife, Spain}
\affil{$^4$Jodrell Bank Centre for Astrophysics, School of Physics and Astronomy, University of Manchester, M13 9PL, UK}
\affil{$^5$Isaac Newton Group of Telescopes, Apart. de Correos 321, 38700 Santa Cruz de la Palma, Spain}
\affil{$^6$Centro de Astrobiologia, CSIC-INTA, Ctra de Torrej\'on a Ajalvir km 4, E-28850 Torrej\'on de Ardoz, Spain}
}

\begin{abstract}
A thorough search of the OGLE-III microlensing project has more than doubled the total sample of PNe known to have close binary central stars. These discoveries have enabled close binary induced morphological trends to be revealed for the first time. Canonical bipolar nebulae, low-ionisation structures and polar outflows are all identified within the sample and are provisionally associated with binarity. We have embarked upon a large photometric monitoring program using the Flemish Mercator telescope to simultaneously test the predictive power of these morphological features and to find more close binaries. Early results are very positive with at least five binaries found so far. This suggests our method is an effective means to expedite the construction of a statistically significant sample of close binary shaped nebulae. Such an authoritative sample will be essential to quantify the degree to which close binary nuclei may shape PNe. 
\end{abstract}
\vspace{-1.25cm}
\section{Introduction}
The pioneering work of H.E. Bond and collaborators (Bond 2000) clearly established the existence of close binary central stars of planetary nebulae (CSPN). 
These binaries evolved through a common-envelope (CE) phase (Iben \& Livio 1993) to typically consist of a low-mass, main-sequence companion that orbits around a hot pre-white dwarf in less than $\sim$1 day. Termination of the CE phase is marked by the ejection of the CE to form a short-lived PN ($t\sim10^4$ years). The presence of a nebula therefore guarantees a post-CE binary is `fresh out of the oven' and ready for use as an ideal probe of the poorly understood post-CE period distribution (e.g. Miszalski et al. 2009a). Furthermore, the same sample is uniquely placed to scrutinise theoretical predictions that the CE phase produces a density contrast responsible for shaping bipolar nebulae (Bond \& Livio 1990; Miszalski et al. 2009b). 

\begin{figure}
   \begin{center}
      \includegraphics[scale=0.65]{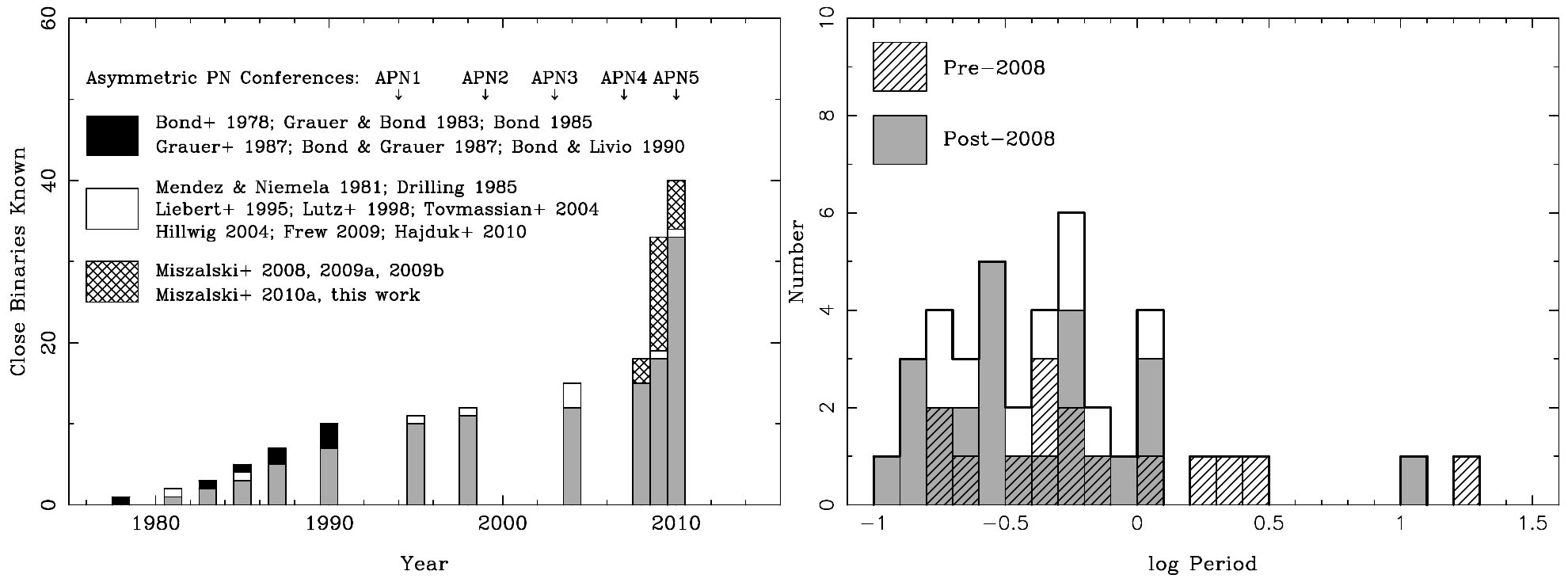}
   \end{center}
   \caption{Discovery timeline of the 40 close binary CSPN known (left) and their orbital period distribution (right). We have added objects in Tab. \ref{tab:newb} and excluded four objects that require further study: PHR 1744$-$3355, PHR 1801$-$2718, PHR 1804$-$2645 and PHR 1804$-$2913 (Miszalski et al. 2009a, 2009b). 
   }

   \label{fig:timeline}
\end{figure}

The discovery of many new close binary CSPN is \emph{fundamental} to develop the full potential of these powerful probes of CE evolution. A meagre 15 close binary CSPN were known before 2008 (Fig. \ref{fig:timeline}) leaving their period distribution and the Bond (2000) 10--15\% binary fraction rather uncertain (De Marco et al. 2008). 
Morphological trends amongst the sample were also somewhat inconclusive with a surprising lack of bipolar morphologies except for NGC 2346 (Bond \& Livio 1990). 
The paucity of close binaries may be explained by limitations in the discovery technique used to find them as well as a general lack of attention to multi-epoch central star studies in the PN literature (the nebula is the focus of most studies).   
All of these binaries were discovered by observing periodic photometric variability caused by irradiation effects, ellipsoidal variation or eclipses. 
The method is straightforward in the absence of bright nebulae and is the only reliable way to find close binaries. Disadvantages include insensitivity to wider binaries (De Marco et al. 2008) and the sometimes slow survey progress due to the quite varied orbital period range of hours to days.

\section{Advances from the OGLE-III Microlensing Survey}

Microlensing surveys are a revolutionary new means to find close binary CSPN. Their extensive time-series photometry and large areal coverage circumvent most limitations of previous searches for close binaries. Of the 15 known binaries only Hf 2-2 was discovered in this way (Lutz et al. 1998). Miszalski et al. (2008b, 2009a) conducted the first detailed study of PNe in any microlensing survey by investigating $\sim$300 Galactic Bulge PNe in OGLE-III (Udalski et al. 2002). Beneficial to this work was the inclusion of faint MASH and MASH-II PNe well-matched to the OGLE-III survey depth (Parker et al. 2006; Miszalski et al 2008a). 
An unprecedented haul of 21 new periodic variables were found which gave a close binary fraction of 17$\pm$5\%. Another major result is the \emph{independent} finding that post-CE population synthesis models overpredict the observed period distribution in agreement with other studies (Rebassa-Mansergas et al. 2008; Davis et al. 2010). 
The vast majority of these new variables have since been spectroscopically confirmed as CSPN (Miszalski et al. 2009b).

\begin{figure}[h]
   \begin{center}
      \includegraphics[scale=0.28]{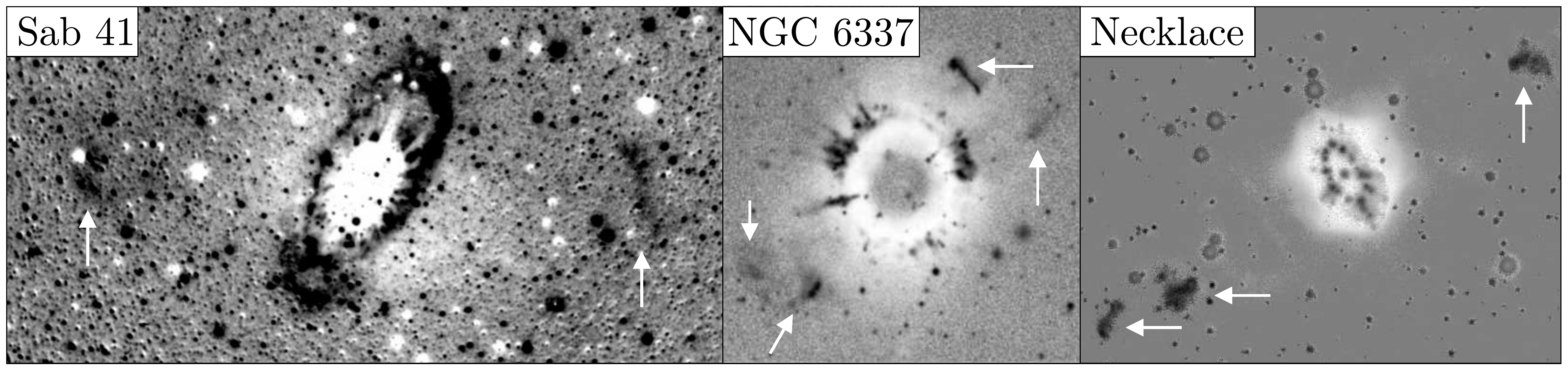}
   \end{center}
   \caption{Sab 41, NGC 6337 and the Necklace all share rings of low-ionisation filaments and presumably polar outflows (arrowed). These features are most evident in [NII] emission (black) against the main nebula bodies in [OIII] (white).}
   \label{fig:jets}
\end{figure}

With the enlarged sample came the opportunity to revisit the previously ambiguous observed morphologies of post-CE PNe. The lack of canonical bipolars as a paradigm was changed by the discovery of the first close binary with $P<1$ day in the canonical bipolar PN M 2-19 (Miszalski et al. 2008b). An implication of this discovery is that the CE phase need not necessarily be avoided to produce a bipolar PN as had been suggested for NGC 2346 (Soker 1997, 1998). With improved images for 30 post-CE PNe Miszalski et al. (2009b) found $\sim$30\% of nebulae had canonical bipolar morphologies. If inclination effects are considered, then this may be as high as 60--70\%, suggesting that CE evolution has a strong preference for producing bipolar nebulae. 
In addition to the underlying bipolar shape, Miszalski et al. (2009b) recognised the elevated presence of poorly understood low-ionisation structures (LIS, Gon{\c c}alves et al. 2001) in $\sim$30\% of the sample and suggested they have a binary origin. Figure \ref{fig:jets} portrays the two main types of LIS: (i) a ring or partial ring of knots or filaments (presumably in the orbital plane), and (ii) the shocked tips of polar outflows or jets (presumably collimated and ejected by the binary e.g. Nordhaus \& Blackman 2006). 

The real test for the trends identified by Miszalski et al. (2009b) comes with additional observations. The high bipolar fraction is supported by L\'opez et al. (2010) who find the general kinematic signature of a torus in many post-CE PNe. Even though the sample size of post-CE PNe is still very small, we have since made further discoveries that strengthen the trends identified by Miszalski et al. (2009b).
Here we describe our work to specifically target PNe befitting these trends for photometric monitoring. In other words, \emph{we are exploring whether we can predict the presence of a close binary based on the nebula morphology alone}.
If confirmed this method offers a way to simultaneously (a) test the morphological trends of Miszalski et al. (2009b), and (b) fast-track the discovery of close binary CSPN!

\section{Prediction of close binarity: The case for a morphologically biased survey}
With the close binary fraction of 17$\pm$5\% now well established by the \emph{morphologically unbiased} survey of Miszalski et al. (2009a), it would be counterproductive to repeat this work. 
Only radically different surveys that target morphologically peculiar sub-samples have the potential to accelerate the construction of a \emph{statistically significant} sample of close binary CSPN. A significant sample could be drawn from either a volume-limited sample (e.g. 40--50 PNe expected amongst 200 PNe within 2kpc, Frew \& Parker 2007) or a much larger magnitude-limited general sample ($\ga$ 200 binaries expected). Building this sample is fundamental to provide the \emph{direct evidence} needed to resolve the debate concerning the shaping of nebula morphologies (Balick \& Frank 2002). 
In the short term the sample may be perceived to be `biased', but as we already have an unbiased survey in Miszalski et al. (2009a), this is an entirely justified sacrifice to make in order to accumulate enough objects suitable for detailed physical study (e.g. Mitchell et al. 2007; Jones et al. 2010) and to search for other trends (e.g. are nebula or torus dimensions correlated with CE ejection efficiency, mass or orbital period?). As a matter of course non-detections will contribute to an important control sample.

\section{Mercator observations}
\begin{figure}
   \begin{center}
      \includegraphics[scale=0.65]{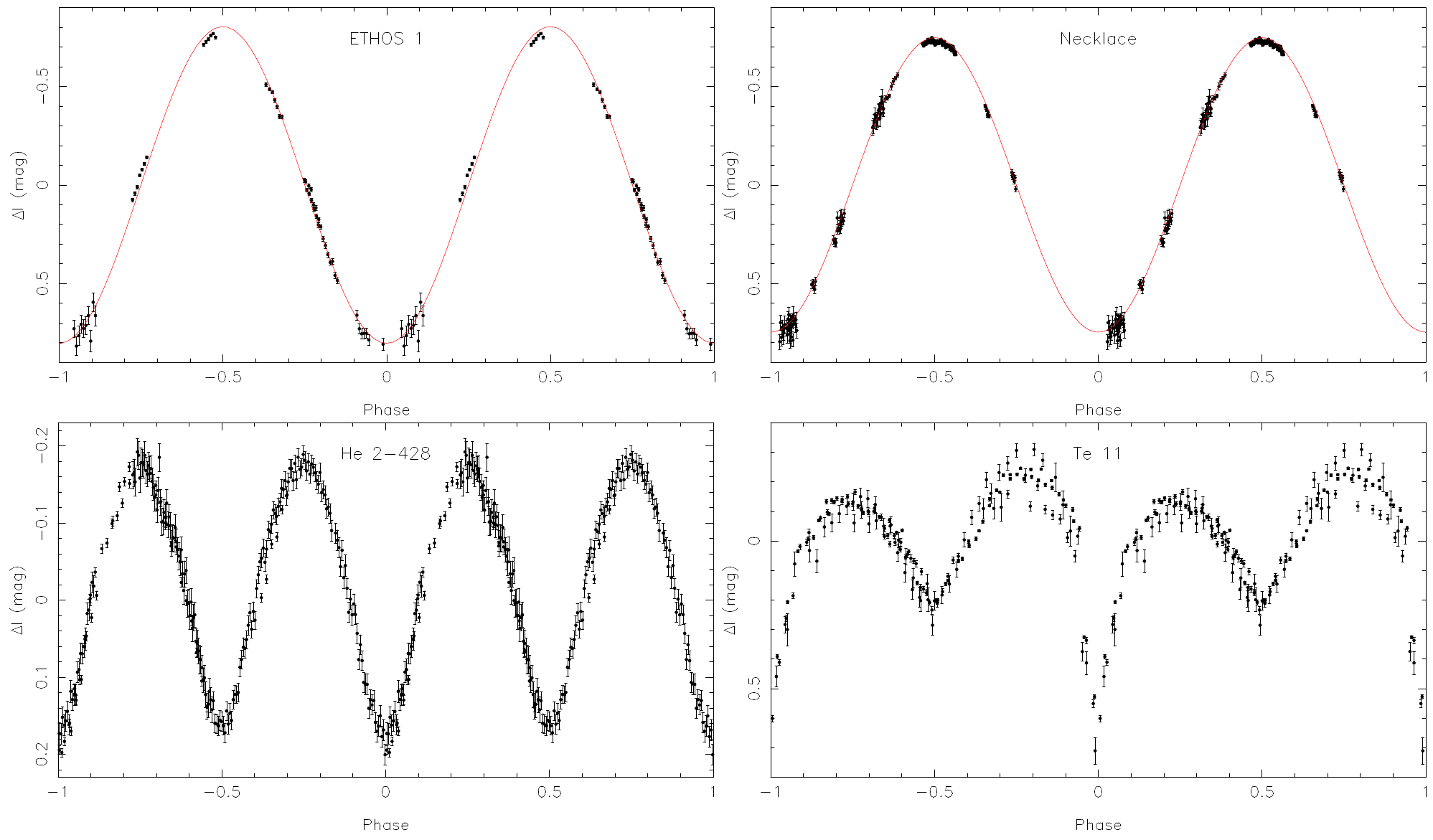}
   \end{center}
   \caption{Lightcurves of the 2009 Mercator discoveries. The Necklace and Te 11 lightcurves include some additional photometry from the Isaac Newton Telescope.}
   \label{fig:lc}
\end{figure}

We have applied our survey strategy to Northern PNe accessible from La Palma with the 1.2-m Flemish Mercator telescope. During an 11 night run starting 24 August 2009 we targeted $\sim$20 PNe that fit the trends identified by Miszalski et al. (2009b). Full details of all objects monitored will be published elsewhere pending analysis of additional 2010 data. The results of our survey were overwhelmingly positive with $\sim$70\% of targets showing variability of some kind, of which five were found to be periodic proving their binary nature. The basic details of these binaries are given in Tab. \ref{tab:newb} and Fig. \ref{fig:lc} shows lightcurves of four from 2009. 
Two of these, the Necklace (Fig. \ref{fig:jets}) and ETHOS 1, were selected for their impressive jet systems. He 2-428 was selected based on its canonical bipolar morphology and suspicions of binarity (Rodr\'iguez et al. 2001), while Te 11 was selected because of its extremely peculiar morphology (Jacoby et al. 2010). A discovery from our 2010 run, NGC 6778, is also mentioned since its underlying bipolar morphology, pair of jets and extensive LIS filaments (Miranda et al. 2010) make it an exemplary supporting case.

\begin{table}
   \centering
   \caption{Recently discovered close binary central stars.}
   \label{tab:newb}
   \begin{tabular}{llrrl}
   \hline\hline
   PN G & Name & Period &  References \\
        &      & (days) & \\
        \hline
        %check PNG
        034.5$-$06.7 & NGC 6778 & 0.15 &  (1), (2)\\
        049.4$+$02.4 & He 2-428 & 0.18 &  (1), (3)\\
        054.2$-$03.4 & Necklace & 1.16 &  (1), (4)\\
        068.1$+$11.0 & ETHOS 1 & 0.53 &  (1), (5)\\
        203.1$-$13.4 & Te 11 & 0.12 &  (1)\\
        221.8$-$04.8 & PHR 0654$-$1045 & 0.63 &  (6)\\
        316.7$-$05.8 & MPA 1508$-$6455 & 12.50 &  (7)\\
        349.3$-$04.2 & Lo 16 & 0.49 &  (8)\\
\hline
   \end{tabular}
   \begin{flushleft}
      (1) this work; (2) Miszalski et al. in prep.; (3) Santander-Garc\'ia et al. 2010; (4) Corradi et al. 2010a, 2010b; (5) Miszalski et al. 2010b; (6) Hajduk et al. 2010; (7) Miszalski et al. 2010a; (8) Frew et al. in prep.
   \end{flushleft}
\end{table}

The most significant aspect of the Necklace, ETHOS 1 and NGC 6778 is the combination of jets \emph{and} a binary nucleus. These outflows were likely ejected as part of CE interactions that in some cases could be dynamo driven (Nordhaus \& Blackman 2006). Emission line ratios of the ETHOS 1 jet tips closely match the FLIERS in NGC 7009 and agree well with predicted ratios of shocks in strongly photoionised environments (Dopita 1997; Raga et al. 2008). Are all FLIERS launched by binaries or could they be produced by other mechanisms?
Only with surveys such as ours can we start to answer this longstanding question (Soker \& Livio 1994).

Also listed in Tab. \ref{tab:newb} are some other pertinent close binary discoveries made recently.
PHR 0654$-$1045 is exceptional since its [WC7] nucleus seems to be the only close binary with a [WR] component (Hajduk et al. 2010). The frequency of close binaries with [WR] components is unknown since the fractions of Hajduk et al. (2010) include very undersampled lightcurves and they did not consider the \emph{indeterminate} binary status of most [WR] sources in OGLE-III as many have very bright nebulae (Miszalski et al. 2009a). We also added MPA 1508$-$6455 from Miszalski et al. (2008a, 2010a) and Lo 16 whose binary status was announced by D. J. Frew at the MASH workshop (Parker 2010). Together with NGC 6778, Lo 16 fits the pattern of Fig. \ref{fig:jets} exceptionally well (Frew \& Parker 2007; Frew et al. in prep.).

\section{Conclusions}
We have begun to test whether the provisional morphological trends identified from a sample of 30 post-CE PNe by Miszalski et al. (2009b) are associated with binarity. These include canonical bipolar nebulae, low-ionisation structures and polar outflows. Photometric monitoring of a biased sample of Northern Galactic PNe with the Flemish Mercator telescope has found at least 5 new close binary CSPN. New PNe surveys are essential to this effort as 3/5  were only discovered in 2009. All have a bipolar morphology, polar outflows or jet systems appear in 3/5, while 2/5 show low-ionisation filaments. These results are most encouraging and suggest with further observations that we can dramatically increase the known population of post-CE PNe in less time than traditional surveys. Only with a statistically significant sample of post-CE PNe can we begin to assemble the direct evidence needed to resolve the debate surrounding the shaping of PNe (Balick \& Frank 2002).

\acknowledgements BM kindly thanks the SOC for an invited talk. 
The team wishes to thank Hans van Winckel for his enthusiastic interest in our project.

\bibliography{miszalski}

\end{document}